\documentclass[conference]{IEEEtran}
\IEEEoverridecommandlockouts

\usepackage{cite}
\usepackage{dblfloatfix}
\usepackage{amsmath,amssymb,amsfonts}
\usepackage{algorithmic}
\usepackage{graphicx}
\usepackage{subfigure}
\usepackage[export]{adjustbox}
\usepackage{textcomp}
\usepackage{xcolor}
\usepackage{booktabs}
\usepackage{vcell}
\usepackage{comment}
\usepackage[hidelinks]{hyperref}
\def\BibTeX{{\rm B\kern-.05em{\sc i\kern-.025em b}\kern-.08em
    T\kern-.1667em\lower.7ex\hbox{E}\kern-.125emX}}
\begin{document}

\title{WebCrowds: An Authoring Tool for Crowd Simulation\\
}

 \author{\IEEEauthorblockN{Gabriel Fonseca Silva}
 \IEEEauthorblockA{\textit{Virtual Humans Lab} \\
 \textit{PUCRS}\\
 Porto Alegre, Brazil \\
 gabriel.fonseca94@edu.pucrs.br}
 \and
 \IEEEauthorblockN{Paulo Knob}
 \IEEEauthorblockA{\textit{Virtual Humans Lab} \\
 \textit{PUCRS}\\
 Porto Alegre, Brazil \\
 paulo.knob@edu.pucrs.br}
 \and
 \IEEEauthorblockN{Rubens Montanha}
 \IEEEauthorblockA{\textit{Virtual Humans Lab} \\
 \textit{PUCRS}\\
 Porto Alegre, Brazil \\
 rubens.montanha@edu.pucrs.br}
 \and
 \IEEEauthorblockN{Soraia Musse}
 \IEEEauthorblockA{\textit{Virtual Humans Lab} \\
 \textit{PUCRS}\\
 Porto Alegre, Brazil \\
 soraia.musse@pucrs.br}
 }

\newcommand\red[1]{{\color{black}#1}}

\maketitle

\begin{abstract}
Crowd simulation is an area of research largely used in the game industry. From the movement of a single NPC to the movement of an entire army, crowd simulation methods can be used to move agents through the environment while avoiding collisions with obstacles and between each other. Thus, it is important that game developers have access to crowd simulation tools that are both powerful and easy to use. In this paper, we present WebCrowds, an authoring tool for crowd simulation which can be used by anyone to build environments and simulate the movement of agents. The results achieved by our research suggest that WebCrowds is easy to use, delivers trustworthy simulation results, and can be used as an authoring tool for game developers who need to simulate crowds in their games.
\end{abstract}

\begin{IEEEkeywords}
Crowd Simulation, Authoring Tool, Virtual Agents, Framework
\end{IEEEkeywords}

\section{Introduction}
\label{sec:introduction}

Several contributions have been made in the area of crowd simulation, given its application in many different fields\footnote{Draft version made for arXiv: \url{https://arxiv.org/}}. Since the pioneer work of Thalmann and Musse~\cite{musse1997model}, many other methods emerged for both microscopic~\cite{pelechano2007controlling} and macroscopic~\cite{hughes2002continuum} points of view and, more recently, combining them both~\cite{silva2019bioclouds}. Ranging from the simulation of a crowd of people or robots~\cite{de2012simulating} to the modeling of personality for simulated virtual agents~\cite{knob2018simulating,durupinar2015psychological}, crowd simulation has a wide range of uses in different fields.
        

Focusing on the area of games, crowd simulation is important for many types of games. Since the movement of Non-Playable Characters (NPCs) to the management of whole armies, crowd simulation methods can be used to move virtual agents from one point to another while avoiding collision with themselves or other obstacles. Suyikno et al.~\cite{suyikno2019feasible} proposed a model to simulate the hiding behavior of NPCs, while Silva et al.~\cite{silva2020moving} proposed a method to move agents forward in time and space fit to be used in "Fog of War" systems. Given the importance of crowd simulation in games, it is essential that game developers have a tool where they can build environments and simulate crowds.

This work presents an authoring tool for crowd simulation. Named as WebCrowds, our tool is divided into two parts: the user Editor, which is accessed through a web Browser, and the crowd simulator, which is served as a web service and is hosted inside a Server. WebCrowds allows users to build environments placing agents, obstacles, and goals in the scene. Users can also load presets from an OBJ file, and even save/load their built scenes using JavaScript Object Notation (JSON) files, to be used for other simulations. \red{Since the simulation is executed on a Server, users can access WebCrowds even with lower-performance computers.} 


\section{Related Work}
\label{sec:related}

\red{This section discusses some work related to authoring tools and techniques for crowd simulation scenarios.} Ulicny et al.~\cite{DBLP:conf/siggraph/UlicnyCT05} presented a way of drawing a crowd simulation where the user could use different types of brushes to create the crowds and change their appearance, behavior, and direction to create a scene to simulate. 
\red{Chen et al. \cite{DBLP:journals/jvca/ChenWL20} presented a way to compose crowd scenarios using natural language processing techniques}, where different words and expressions represent the attributes and behavior of each agent. Some of these expressions may contain emotions, such as "walk happily" and "walk sadly". These agents are created using data provided by a user, and this data is used to extract character keywords. \red{This model was also employed by Liu et al.~\cite{DBLP:journals/jvca/LiuWC20}, where keywords such as quantity, character type, behavioral verb, location, and object, are taken into account.} The data structure of the crowd scene is divided into three components: crowd, behavior, and destination. Each of these components has its keywords. 

\red{Colas et al.~\cite{sketchingFields2022} presented a model for authoring crowd simulations using sketches. The user draws curves in the interface, which are used to compose an interaction field grid, influencing the steering behavior of agents. Other agents and obstacles in the scene are also taken into account when composing the grid, which allows agents to avoid collision with one another or move behind obstacles. Kim and Sung~\cite{skecthingAR2022} presented a model using sketching in combination with augmented reality, where users can use dry-erase markers to compose scenes. The authors use computer vision techniques to identify shapes and colors in the sketches to define which type of object (e.g. goals, paths, and trees) should be placed in that position.
}

\red{The work found in the literature focuses on crowd animation, where users may define behaviors or directly input an animation for virtual characters; or explore new user interaction techniques, such as sketching, augmented reality, and natural language. Differing from them, WebCrowds provides information on the constructed environment and the simulation run, allowing users to evaluate trajectories and densities of virtual agents and make any adjustments desired.}

\section{Proposed Model}
\label{sec:model}

In this section, we present WebCrowds, an authoring tool for crowd simulation which can be used by anyone to build environments and simulate the movement of virtual agents. Next, we present the main elements of our model.

\subsection{System Architecture}
\label{sec:system_architecture}

\begin{figure*}[!htbp]
    \centering
    \includegraphics[width=0.9\textwidth]{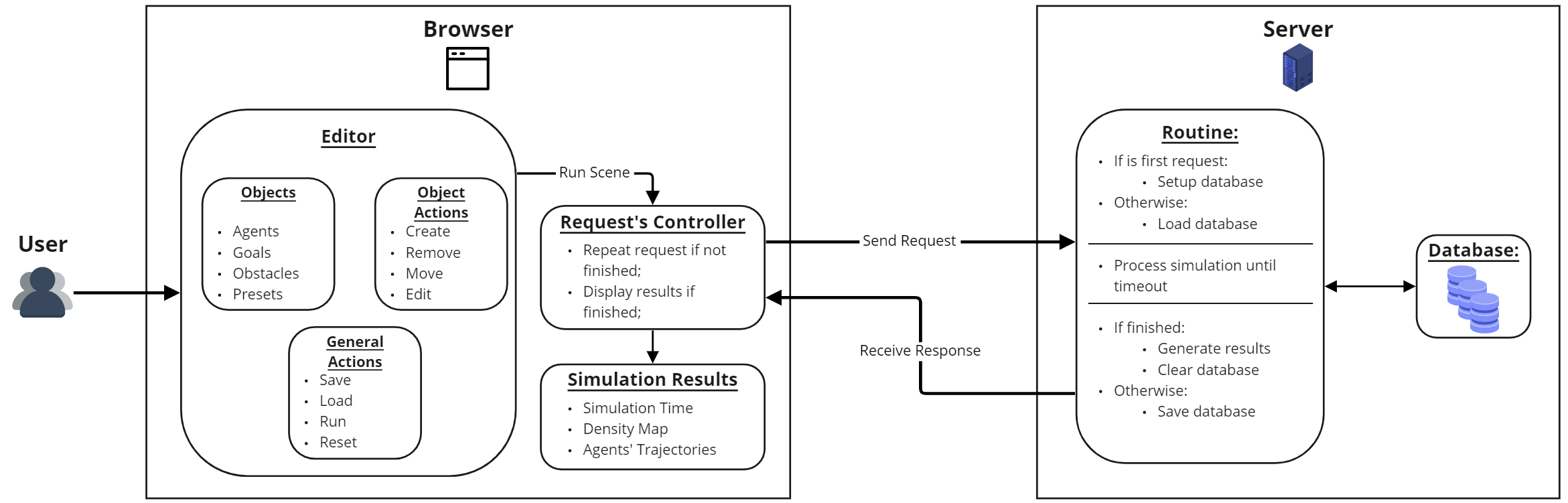}
    \caption{
    Pipeline of WebCrowds. Users can use the Editor through a web Browser. Requests for a given simulation are sent to the Server. The Server returns the results when the simulation is finished, which are shown in the Browser.}
    \label{fig:overview}
\end{figure*}

Figure~\ref{fig:overview} presents an overview of our proposed model, which is available online and can be accessed through a Browser\footnote{WebCrowds is available at: \url{ https://vhlab.com.br/projects/webcrowds/}.}. Inside WebCrowds, the users can create an environment by adding, positioning, and editing objects. In the bottom right of Figure~\ref{fig:defaultUI}, highlighted in purple, it is possible to see the Objects that can be used, namely: Agents, Goals, Obstacles, and Presets. More details about these objects are going to be provided in Section~\ref{sec:editor}. 

Following the pipeline presented in Figure~\ref{fig:overview}, when the user opens WebCrowds in their Browser, they have access to the Editor, where the Scene can be edited to be then simulated. Once the agents, goals, and obstacles are included, the user can run the simulation. When the user runs the simulation, the Request Controller gathers information about the Objects in the Scene (e.g. their positions and properties) and sends a request to the Server (Send Request). In its turn, when the Server receives a new simulation, it triggers its own Routine. In short, the Server executes the simulation with the parameters sent from the Browser and, once it is finished, generates the results. More details about the Server are provided in Section~\ref{sec:server}.

The results are generated and sent back to the Request Controller (Receive Response), which will display them in the browser (Simulation Results). These results are comprised of three metrics: a Density Map (which shows the density of agents during the simulation), the Agent's Trajectories (which shows the trajectories of each agent), and the Simulation Time. Finally, In Figure~\ref{fig:overview}, it is possible to see that the Server (on the right) also connects with a Database. It was done to overcome request timeout issues, which would prevent the simulation from finishing. 

\subsection{Editor}
\label{sec:editor}

In this section, we describe the three main elements of WebCrowds: Objects, Object Actions, and General Actions. All three can be seen in Figure~\ref{fig:defaultUI}, highlighted in purple (bottom right), yellow (bottom left), and blue (top), respectively.

\subsubsection{Objects}
\label{sec:system_objects}

In WebCrowds, the users can deal with four types of Objects: Agents, Goals, Obstacles, and Presets. All of them can be seen in the bottom right of Figure~\ref{fig:defaultUI}, highlighted in purple. These objects can be positioned within the scene's environment and are detailed next:

\begin{itemize}
    \item Agents: Allows the creation of agents spawn areas, represented as blue squares in the central region of Figure~\ref{fig:defaultUI};
    \item Goals: Allows the creation of goals for the agents to achieve, represented as green circles in Figure~\ref{fig:defaultUI};
    \item Obstacles: Allows the creation of obstacles in the environment that should be avoided by the agents, represented as red squares in Figure~\ref{fig:defaultUI}; and
    \item Presets: Allows the user to load built-in scenarios into the environment. Presets behave as a set of Obstacles that can't be modified. We modeled five possible presets, as presented in Figure~\ref{fig:presetsUI}.
\end{itemize}

\subsubsection{Object Actions}
\label{sec:system_object_action}

In WebCrowds Object Actions are simply the actions that users can perform with Objects. The users can control four types of Objects Actions: Create, Remove, Move and Edit. All of them can be seen in the bottom left of Figure~\ref{fig:defaultUI}, highlighted in yellow, and are detailed as follows:

\begin{itemize}
    \item Create: Allows the creation of any type of Object (i.e. Agent, Obstacle, Goal, or Preset);
    \item Remove: Allows the removal of Objects present in the environment;
    \item Move: Allows the user to move any Object present in the environment, positioning it where he/she sees fit; and
    \item Edit: Allows the user to edit any Object present in the environment. The editable options depend on the type of Object. When editing an Agents object, users can change the number of virtual agents to be spawned in that area, as well as the Goal that these agents want to reach. When editing an Obstacle, the user can change its size and rotation. Goals and Presets have no editing options.
\end{itemize}

\subsubsection{General Actions}
\label{sec:system_general_action}

\begin{figure}[!htb]
  \centering
  \subfigure[fig:defaultUI][Editor's interface. General Actions are available at the top of the UI, highlighted in cyan; Object Action at the bottom-left, highlighted in yellow; and Objects at the bottom-right, highlighted in purple.]{\includegraphics[width=0.4\textwidth]{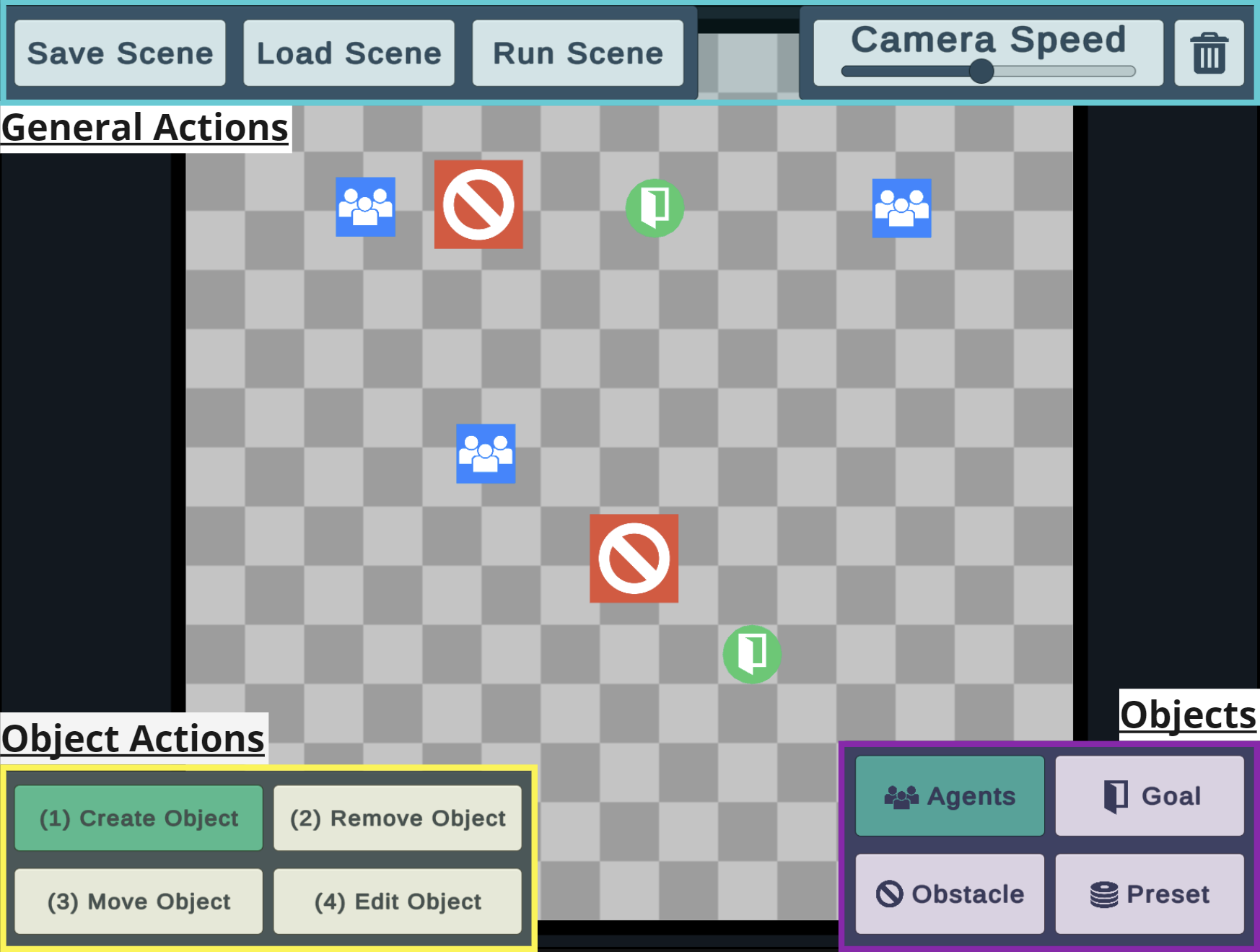}
  \label{fig:defaultUI}}
  \subfigure[fig:presets][Preset selection screen. Five built-in presets are available, which behave as unmodifiable Obstacles.]{\includegraphics[width=0.4\textwidth]{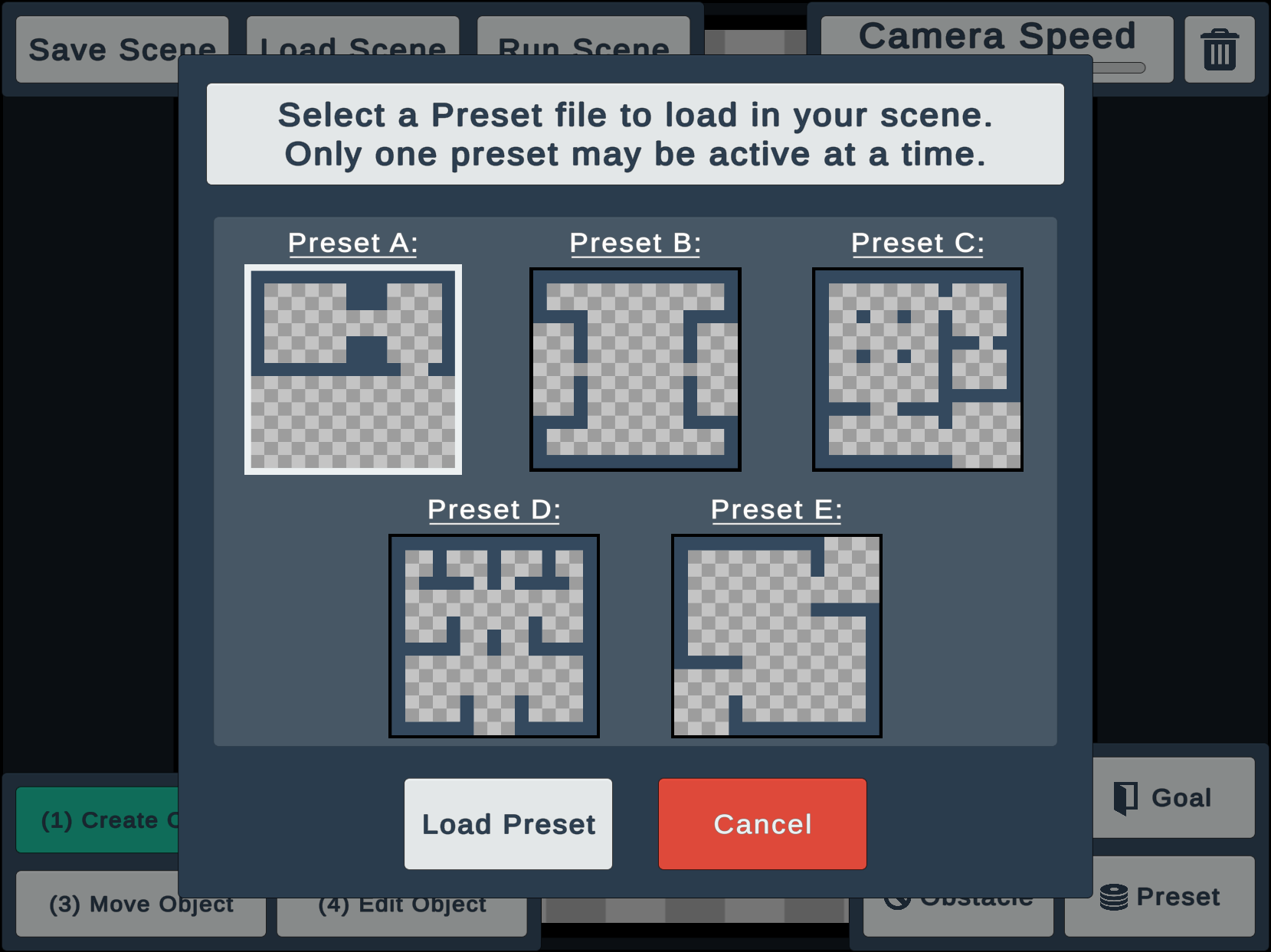}
  \label{fig:presetsUI}}
    \caption{Editor's User Interface. Objects, Object Actions, and General Actions are presented in (a). The preset selection screen is presented in (b).}
    \label{fig:editorUI}
\end{figure}

In WebCrowds we have five General Actions: Save Scene, Load Scene, Run Scene, Camera Control and Reset Scene, as next detailed. All of them can be seen at the top of Figure~\ref{fig:editorUI}, highlighted in blue, where the Recycle Bin icon represents the Reset action.

\begin{itemize}
    \item Save Scene: Allows the user to save the current environment into her/his computer/device. All the Objects present in the environment are saved into a JSON file and can be used in later simulations;
    \item Load Scene: Allows the loading of a JSON file from the user's computer/device. Ideally, this JSON file is the same one that was generated with the Save Scene action, but any JSON file can be used as long as it follows the same information pattern;
    \item Run Scene: Sends a request to the Server to run a simulation with the environment built;
    \item Camera Control: Allows the user to perform some actions concerning the camera, such as Movement (with WASD keyboard keys), changing the Camera Speed (as shown in Figure~\ref{fig:defaultUI}) and Zoom in/Zoom out (using the mouse wheel); and
    \item Reset Scene: Allows the user to clear the environment and start a new one, being accessed by the Recycle Bin icon in Figure~\ref{fig:defaultUI} (top-right).
\end{itemize}

\subsection{Server}
\label{sec:server}

In WebCrowds, the Server is responsible for running the simulation defined in the Editor, generating the Results, and sending them back to the Browser. The simulation runs using the BioCrowds simulation model~\cite{de2012simulating}, developed in Python. However, since the Server is decoupled from the Editor, any crowd simulation method can be applied. Moreover, we also added a path planning algorithm (i.e. A*), so agents can easily find their paths. \red{In our current implementation, the Database stores the position and size of all Objects, in addition to each agent's goal and trajectory.}

\begin{figure}[!htb]
  \centering
  \subfigure[fig:agentTrajectory][Agents' Trajectories.]{\includegraphics[width=0.4\textwidth]{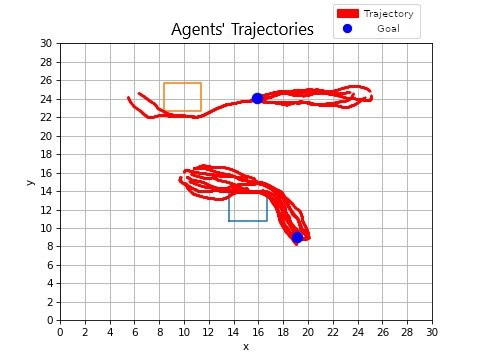}
  \label{fig:output_trajectories}}
  \subfigure[fig:densityMap][Density Map.]{\includegraphics[width=0.4\textwidth]{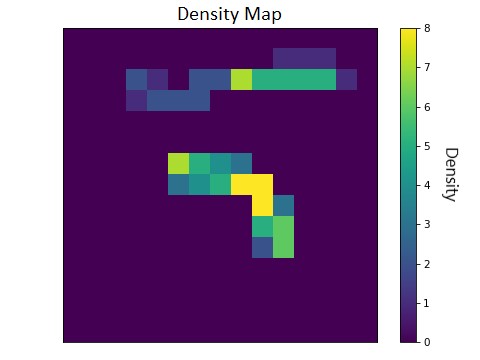}
  \label{fig:output_density}}
    \caption{
    Results generated by the Server. Agents' Trajectories~\ref{fig:output_trajectories} show the trajectories of all agents of the simulation, while Density Map~\ref{fig:output_density} shows the concentration of agents in the environment.}
    \label{fig:graphAndOutline}
\end{figure}

As commented in the System Architecture (Section~\ref{sec:system_architecture}), the results generated by the Server are comprised of a Density Map, Agents' Trajectory, and the total Simulation Time. Figure~\ref{fig:graphAndOutline} presents an example with Agents' Trajectories and Density Map figures generated by the Server. 
In Figure~\ref{fig:output_trajectories}, it is possible to see the two obstacles (orange and blue squares) that affect the trajectories of the agents. Goals are represented as the blue points, and the trajectories are represented as the red lines, one line for each agent.

Figure~\ref{fig:output_density} presents an example of a Density Map, generated with the same simulation which generated the trajectories from Figure~\ref{fig:output_trajectories}. The color density scale is presented on the right as a vertical color bar. It ranges from deep blue to yellow, with deep blue denoting the spaces where no agents were present and yellow the spaces with the highest concentration of agents. Finally, we hosted our Server routine inside a Heroku free account.

\section{User Experiment}  
\label{sec:experiment}

A user experiment was designed to evaluate two main aspects of WebCrowds: (a) "Is the system easy to use?", and (b) "Are the simulation results clear?". Users were presented with an online questionnaire (described in Table~\ref{tbl:statements}), where they were asked to open WebCrowds, perform tasks (e.g., creating a simulation scenario and observing its results), and answer questions related to these tasks. Participants \red{over 18 years old} of all backgrounds and technical knowledge were invited\red{, as volunteers, via social media} to perform the experiment and the entire process was made without the supervision of the authors. The estimated duration of the experiment was 35 to 45 minutes.

Experts with knowledge of real-world public safety simulation scenarios (i.e., fire prevention and evacuation) also participated in the experiment. Experts gave their consent to being recorded with screen capture and a webcam. 
The goal was to evaluate the interface with application experts.
After completing the questionnaire, experts were interviewed and asked to comment on possible features to employ WebCrowds as an assisting tool for public safety professionals.

For this experiment, we defined some constraints. The simulation environment was set to a default size of 30 by 30 meters, each Agents object was limited to a maximum of 10 agents, and the entire Scene was limited to 100 agents total. Obstacles were limited to a minimum size of 2 by 2 meters, and maximum of 20 by 20.

\subsection{Questionnaire}
\label{sec:questionnaire}

The online questionnaire starts with a brief description of the project, the estimated duration, and the author's contact information. Following ethical guidelines for applying the questionnaire, all participants were asked if they agreed to grant access to their answers and personal information regarding age, gender, educational level, occupation, and prior experience with technology and simulation tools\footnote{Project Estudos e Avaliações da Percepção Humana em Personagens e Multidões Virtuais, number 46571721.6.0000.5336, approved by the Ethics Committee of Pontifical Catholic University of Rio Grande do Sul.}. Given their agreement, participants were presented with instructions to access WebCrowds.

The main portion of the questionnaire is divided into 7 steps. Each step is composed of: a list of tasks to be performed in WebCrowds; statements regarding the performed tasks. The steps are detailed next: 
\begin{itemize}
    \item Step 1: Participants are free to interact with WebCrowds before performing tasks;
    \item Step 2: Participants should create one Agents and a Goal object. Set the Goal of the Agents and run the simulation;
    \item Step 3: Participants should create an Obstacle between the objects of the previous step. Run the simulation;
    \item Step 4: Participants should create any objects as desired and edit the properties of Agents and Obstacles. Run the simulation;
    \item Step 5: Participants should load a Preset and modify the scene as desired. Run the simulation;
    \item Step 6: Participants should save the Scene to a file, then, Clear the Scene. Open the saved Scene and run the simulation; and
    \item Step 7: Participants were asked to provide their final feedback on WebCrowds.
\end{itemize}

Table~\ref{tbl:statements} presents the statements given to participants at each Step, for their evaluation.
Statements regarding the system being easy to use are marked with a ``*", and statements regarding WebCrowds presenting clear results are marked with a ``º".
For each statement, participants gave their feedback using a five-level Likert scale from Strongly Disagree (1) to Strongly Agree (5). 
From Step 2 until Step 6, it was also included the following questions that could be answered with textual information:
\begin{itemize}
    \item ``If you had any difficulty performing this step, please describe it": an optional text field;
    \item ``If you have any suggestions or comments regarding the tasks of this step, please share": an optional text field; and
    \item ``Do you wish to continue to the next steps?": a Yes or No question, so the participant may leave the experiment at any moment.
\end{itemize}

\begin{table}[ht]
\centering
\caption{Table of statements presented to participants at each Stage of the questionnaire.}
\label{tbl:statements}
\begin{tabular}{@{}lcl@{}}
\toprule
\multicolumn{1}{c}{Step}                                               & ID                                                      & Statement                                                                                                                            \\ \midrule
\multicolumn{1}{c}{1}                                                  & ------                                                  & No statements in this Step.                                                                                                          \\ \cmidrule(l){2-3} 
\multicolumn{1}{c}{2}                                                  & S2-A*                                                  & Creating new objects is a simple process;                                                                                            \\
                                                                       & S2-B*                                                  & Setting the Goal of an Agents object is a simple process.                                                                            \\
                                                                       & S2-Cº                                                  & The information on the Agents' Trajectory is clear.                                                                                  \\ \cmidrule(l){2-3} 
\multicolumn{1}{c}{3}                                                  & S3-A*                                                  & Creating a new Obstacle is a simple process.                                                                                         \\
                                                                       & \begin{tabular}[c]{@{}c@{}}S3-Bº\\ \newline\end{tabular} & \begin{tabular}[c]{@{}l@{}}The information on the Simulation Time has been\\ impacted by the added Obstacle.\end{tabular}            \\
                                                                       & \begin{tabular}[c]{@{}c@{}}S3-Cº\\ \newline\end{tabular} & \begin{tabular}[c]{@{}l@{}}The information on the Agents' Trajectory is clear\\ and consistent with the added Obstacle.\end{tabular} \\ \cmidrule(l){2-3} 
\multicolumn{1}{c}{4}                                                  & S4-A*                                                  & Editing an object's properties is a simple process.                                                                                  \\
                                                                       & \begin{tabular}[c]{@{}c@{}}S4-Bº\\ \newline\end{tabular} & \begin{tabular}[c]{@{}l@{}}The information on the Agents' Trajectory\\ is consistent with the built environment.\end{tabular}        \\
                                                                       & \begin{tabular}[c]{@{}c@{}}S4-Cº\\ \newline\end{tabular} & \begin{tabular}[c]{@{}l@{}}The information on the Density Map clearly\\ indicates the areas most used by agents.\end{tabular}        \\ \cmidrule(l){2-3} 
\multicolumn{1}{c}{5}                                                  & S5-A*                                                  & Creating a Preset object is a simple process.                                                                                        \\
                                                                       & \begin{tabular}[c]{@{}c@{}}S5-Bº\\ \newline\end{tabular} & \begin{tabular}[c]{@{}l@{}}The information on the Agents' Trajectory\\ is consistent with the built environment.\end{tabular}        \\
                                                                       & \begin{tabular}[c]{@{}c@{}}S5-Cº\\ \newline\end{tabular} & \begin{tabular}[c]{@{}l@{}}The information on the Density Map\\ clearly indicates the areas most used by agents.\end{tabular}        \\ \cmidrule(l){2-3} 
\multicolumn{1}{c}{\begin{tabular}[c]{@{}c@{}}6\\ \newline\end{tabular}} & \begin{tabular}[c]{@{}c@{}}S6-Aº\\ \newline\end{tabular} & \begin{tabular}[c]{@{}l@{}}The results of this simulation are similar to the\\ results of the previous simulation.\end{tabular}      \\ \cmidrule(l){2-3} 
\multicolumn{1}{c}{7}                                                  & S7-A*                                                  & The tool is easy to use.                                                                                                             \\
                                                                       & S7-Bº                                                  & The results presented by the simulations are clear.                                                                                  \\\midrule
                                                                 \multicolumn{3}{l}{*Statements regarding a) Easy to Use.}                                       \\
                                                                     \multicolumn{3}{l}{ºStatements regarding b) Clear Results.}          
                                                                       \\\bottomrule
\end{tabular}
\end{table}

\section{Results and Discussion}
\label{sec:results}

\begin{figure}[!htb]
  \centering
  \subfigure[]
  {\includegraphics[width=0.2\textwidth]{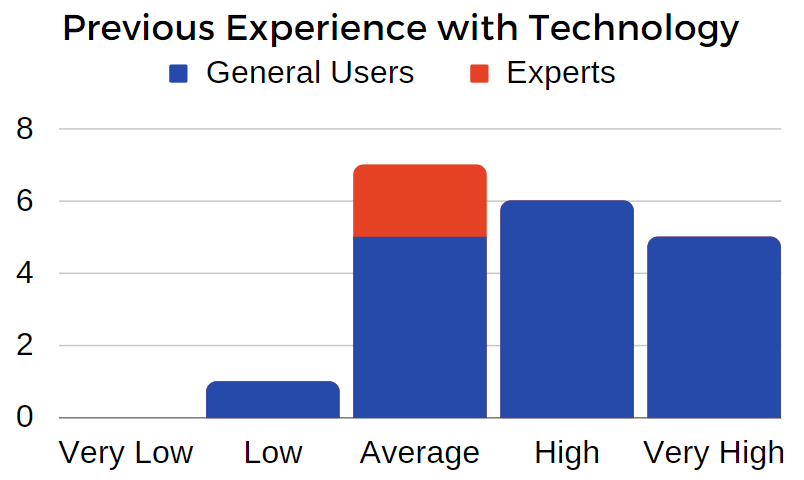}
  \label{fig:previous_experience_technology}}
  \subfigure[]
  {\includegraphics[width=0.2\textwidth]{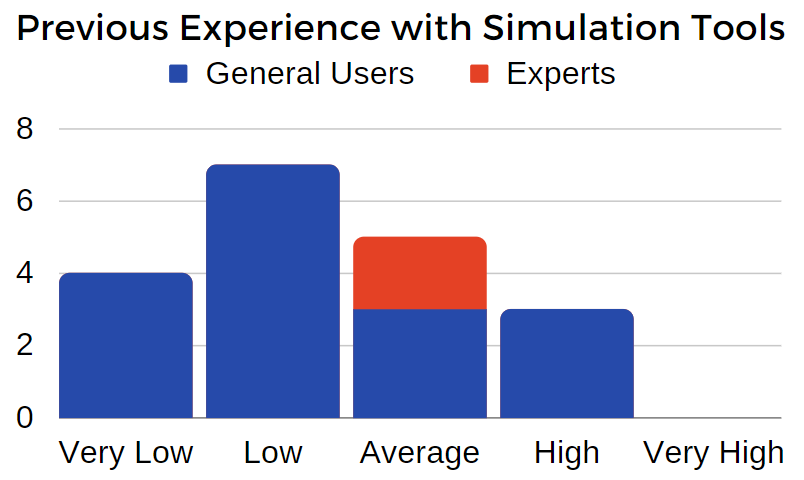}
  \label{fig:previous_experience_simulation}}
    \caption{Participants' previous experience with technology and simulation tools.}
    \label{fig:previous_experience}
\end{figure}

The experiment had a total of 20 participants, being it 18 general users and 2 experts. Regarding the age of the participants: 3 participants (15\%) were 18 to 20 years old; 11 (55\%) were 21 to 29 years old; 2 (10\%) were 30 to 39 years old; 2 (10\%) were 40 to 49 years old; 1 (5\%) was 50 to 59 years old; and 1 (5\%) was over 60 years old. Regarding gender: 11 participants were male (55\%); 8 were female (40\%); and 1 (5\%) preferred to not self-identify.
Participants had varying degrees of experience with technology and simulation tools, as presented in Figure~\ref{fig:previous_experience}. Although the majority of participants considered themselves as having average or higher experience with technology, most of them had little or no prior contact with simulation tools. 
Both expert users considered themselves as having an average experience with both technology and simulation tools.

Table~\ref{tbl:answers} presents the average and standard deviations for the answers of both general users and experts, concerning all statements of the questionnaire. It is important to remember that the Likert Scale varied from 1 to 5, thus, the maximum possible Average is 5.
It is possible to notice that the worst average was scored by general users concerning S2-B (``Setting the Goal of an Agent's object is a simple process''). 
Indeed, looking at users' feedback, this sub-task was mentioned as the most frequent issue by the users: from the 20 participants, 7 commented about this topic. More details about the feedback of the users are provided in Section~\ref{sec:user-feedback}. Moreover, the highest standard deviation was observed with general users in S5-B (``The information on the Agents’ Trajectory is consistent with the built environment''). It suggests that, although having a high average score (4.44), the results presented to the users may have caused a little confusion. One of the participants suggested that both maps (Density and Trajectories) could have more legends and other types of indications. Finally, S3-A (``Creating a new Obstacle is a simple process'') presented the highest average score (4.833) and lowest standard deviation (0.383) for general users, which suggests that this sub-task was the one that general users felt more comfortable with.

Concerning the experts' evaluation, all the answers lay between 4 and 5 on the Likert scale. 
Suggestions were also made and are presented in the next section. Moreover, we can compare the results achieved by general users and experts. In order to investigate if the answers vary too much between general users and experts, we rely on the ANOVA test. The results reveal a uniformity in the participants' answers (F(4.19)=3.34,p=.07)), which suggest that both general users and experts had similar impressions about WebCrowds. It may also suggest that WebCrowds is, despite some limitations, simple enough for general users to use and powerful enough to be used as a simulation tool by experts of different areas. 

\begin{table}[ht]
\centering
\caption{Average and Standard Deviation of general users for all statements of the questionnaire. Lower values are underlined while the higher values are in bold.}
\begin{tabular}{@{}ccc@{}}
\toprule
                  & \multicolumn{2}{c}{General Users (18)}                                \\ \cmidrule(l){2-3} 
ID                & Average                           & Std Dev                           \\ \midrule
S2-A*             & 4.55                              & 0.78                              \\
S2-B*             & \underline{3.44} & 0.78                              \\
S2-Cº             & 4.55                              & 0.78                              \\
S3-A*             & \textbf{4.83}    & \underline{0.38} \\
S3-Bº             & 4.55                              & 0.78                              \\
S3-Cº             & 4.72                              & 0.75                              \\
S4-A*             & 4.38                              & 0.69                              \\
S4-Bº             & 4.72                              & 0.57                              \\
S4-Cº             & 4.50                              & 0.78                              \\
S5-A*             & 4.50                              & 0.78                              \\
S5-Bº             & 4.44                              & \textbf{1.04}    \\
S5-Cº             & 4.77                              & 0.42                              \\
S6-Aº             & 4.50                              & 0.98                              \\
S7-A*             & 4.33                              & 0.59                              \\
S7-Bº             & 4.66                              & 0.48                              \\ \midrule
a) Easy to Use*   & 4.34                              & 0.87                              \\
b) Clear Resultsº & 4.60                              & 0.75                              \\ \bottomrule
\end{tabular}
\label{tbl:answers}
\end{table}


\subsection{User Feedback}
\label{sec:user-feedback}

As described in Section~\ref{sec:questionnaire}, participants were free to provide feedback and describe any difficulty they had when performing the tasks of a step.
The most frequent issue mentioned was regarding the task of setting a Goal for an Agents object in Step 2. A total of 7 participants commented on this topic, describing that the procedure could be streamlined by: having Agents select a default Goal in the environment, which could later be edited by users; by allowing objects to be edited in groups; and by giving clearer visual feedback that a Goal is set to a given Agents object. The difficulty found in the task is also apparent for participants that did not provide feedback, given that the statement S2-B has the lowest average score and highest standard deviation in Table~\ref{tbl:answers}. Other suggestions were the inclusion of an interactive tutorial for first-time users, undo-redo options, and combining the Move and Edit Object Actions.

Concerning the comments made by the experts, they commented on two main topics that they consider when designing and evaluating evacuation plans: the usage of the space, related to the type of building (e.g. restaurant, shopping) and its population; and units of passage, related to size and quantity of exits. Large buildings with many people inside would require a higher number of units of passage to increase the flow of people and achieve a certain evacuation time, as defined by law. They also commented that additional signalization (e.g. exit signs) is required if the maximum distance traveled during the evacuation surpasses a given threshold. In order to employ our proposed model as an assisting tool for these professionals, WebCrowds could provide an average and maximum distance traveled by agents and allow the editing of the size of Goals and usage of spaces within the Editor's environment.
The inclusion of another type of Object (i.e. Sign) could also be provided, allowing the user to include signalization in the environment.

\section{Final Considerations}
\label{sec:finalConsiderations}

This paper presented WebCrowds, an authoring tool for crowd simulation that can be used by anyone to build environments and simulate the movement of agents. WebCrowds is divided into the Editor and the Server. The users have access to the Editor, where they can freely build their own scene, run simulations and receive the results. We conducted experiments with users, both general and experts. The results of such experiments suggest that WebCrowds was easy to use, delivered comprehensible simulation results back to the user, and can be a valuable authoring tool for people to simulate the movement of crowds. It can be integrated with game engines in two ways. First, the JSON files generated by the editor contain the information of every object and can be parsed to produce elements in a game level/scene. 
Second, the engine may produce the information required to run a WebCrowds simulation, so developers can evaluate the possible trajectories and densities of the agents in the environment they are building.

WebCrowds has some limitations. 
There are already many extensions for BioCrowds that could be used to enhance the behavior of the agents, like personality~\cite{knob2018simulating} and emotion contagion~\cite{borges2017giving}. 
Moreover, 
A* can make agents choose to squeeze themselves in crowded places instead of looking for paths with lower densities. One possible solution would be to use a dynamic path-planner (e.g. D*), which would be able to change agents' paths during the simulation.


In future work, there are some avenues to explore. \red{Further testing can be made with game developers, exploring applications such as level design and path-planning for virtual characters.} Besides the metrics presented in this work as results of the simulation (i.e. Density Map, Trajectories Map, and Simulation Time), we could implement other metrics, such as average and maximum distance traveled for agents. Moreover, identified limitations could be tackled and mitigated, like agents' behavior enhancing and dynamic path-planning. Finally, we can use the feedback provided by the users to improve WebCrowds. For instance, defining default Goals for Agents, defining size for Goals, and allowing the edit of groups of objects.


\section*{Acknowledgments}
This work is partially supported by CNPq and CAPES.

\bibliographystyle{IEEEtran}
\bibliography{bib}



\end{document}